\documentclass[12pt,titlepage]{article}
\usepackage{graphicx}
\usepackage{amssymb}

\font\smallish=cmr8 scaled \magstep1
\font\grande=cmr10 scaled \magstep4
\font\medio=cmr10 scaled \magstep2
\outer\def\beginsection#1\par{\medbreak\bigskip
      \message{#1}\leftline{\bf#1}\nobreak\medskip
\vskip-\parskip
      \noindent}

\def\laq{\raise 0.4ex\hbox{$<$}\kern -0.8em\lower 0.62
ex\hbox{$\sim$}}
\def\gaq{\raise 0.4ex\hbox{$>$}\kern -0.7em\lower 0.62
ex\hbox{$\sim$}}
\def\beq{\begin{equation}}
\def\eeq{\end{equation}}
\def\bea{\begin{eqnarray}}
\def\eea{\end{eqnarray}}
\def\bean{\begin{eqnarray*}}
\def\eean{\end{eqnarray*}}

\begin{document}
\bibliographystyle {unsrt}
\titlepage
\begin{flushright}
\vspace{15mm}
CERN-PH-TH/2004-192 \\
hep-th/0410166 \\

\end{flushright}
\vspace{15mm}
\begin{center}
{\grande String-theoretic unitary S-matrix at the threshold of black-hole production}\\

\vspace{15mm}

\vspace{6mm}
 \large{ G. Veneziano}
\vspace{6mm}

{\sl Theory Division, CERN, CH-1211 Geneva 23, Switzerland} 

{\sl and} 

{\sl Coll\`ege de France, 11 place M. Berthelot, 75005 Paris, France} 

\end{center}

\vskip 2cm
\centerline{\medio  Abstract}
\vskip 5mm
\noindent
  Previous results  on trans-Planckian  collisions  in  superstring theory are  rewritten in terms of an explicitly unitary S-matrix whose range of validity covers a large region of the energy impact-parameter plane. Amusingly, as part of this region's border is approached,  properties of the final state start resembling those expected from the evaporation of a black hole  even well below its production threshold.  More specifically,  we conjecture that, in an energy window extending up to such a threshold,  inclusive cross sections satisfy a peculiar ``antiscaling" behaviour, seemingly preparing for a smooth transition to black-hole physics. 

\vspace{5mm}

\vfill
\begin{flushleft}
CERN-PH-TH/2004-192 \\
October 2004\\
\end{flushleft}

\newpage
\section{Introduction}

Since Hawking's theoretical discovery that black holes emit thermal radiation \cite{Haw},  the issue of a possible loss of information/quantum coherence in processes where a black hole (BH) is produced and then evaporates has been the subject of much debate.
Progress coming from string theory, in particular on the microscopic understanding of black-hole entropy \cite{entropy}  and on 
the AdS-CFT correspondence \cite{AdS}, have lent strong support \cite{MBR}  to the ``no-loss" camp.  Hawking himself appears to have conceded this point \cite{HawD} by presenting his own arguments in favour of preservation of quantum coherence, based on a topological distinction between metrics that enter (through a Euclidean path integral)  the quantum description of the
scattering process, as opposed to those describing an eternal black hole.

Even if one may consider the information-paradox to be conceptually solved, several issues still remain unclear. One would like to understand, for instance, how exactly information is retrieved and what this implies on the
properties of the (fully coherent) final state that a given pure initial state generates through its unitary evolution.
A few ideas on this issue have been floating around: at one extreme, it has been suggested \cite{Amati} that black-hole formation and decay simply do not occur for a pure initial quantum state and would only emerge after tracing over initial states (i.e. if one starts already from a density matrix). Less radical solutions \cite{solutions} suggest that information is recovered during the whole evaporation process, in subtle correlations among the final particles or that it is given back at the very end through some black-hole ``remnants". In general, one may suspect that the fate of the classical (space-like) singularity lurking inside the black-hole horizon and/or the fate of the singularity encountered at the end of the evaporation process in a finite theory of quantum gravity, such as superstring theory, should have a bearing on the issue of how information is fully recovered at late-enough times.

In this paper we will not be able to give a definite answer to this problem. However, using the fact that, in string theory,
the threshold for BH formation, $E_{\rm th}$,  can be made arbitrarily high (in string or Planck units) by taking a sufficiently small string coupling, we will analyse the unitary S-matrix that describes the collision process in a  large energy interval up to such a threshold. We will thus be able to analyse the detailed structure of the final state and see to what extent its properties,  as $E_{\rm th}$ is approached,  resemble more and more those predicted by the Hawking process. To our satisfaction, the matching near $E_{\rm th}$ turns out to be very smooth. This is almost certainly related to the well-known correspondence \cite{Corr1} between strings and black holes at  $M \sim E_{\rm th}$, point at which 
 the Hawking temperature coincides with the Hagedorn temperature of string theory,  back-hole and string entropy  go smoothly into one another, and so do many other physical quantities \cite{Corr2}. Our findings  give further support to the belief that quantum coherence is fully recovered even when the region of ``large" semi-classical black holes is reached.

There is a second, less fundamental, but phenomenologically  interesting reason for studying high-energy collisions in the region $M_{\rm P} < E < E_{\rm th}$.  Models have been proposed \cite{LED}  in which gravity, being sensitive to some ``large" extra dimensions of space, becomes strong at an effective energy scale $M_D$ that is much much smaller than the ``phenomenological" $4$-dimensional Planck energy $M_{\rm P} \sim 10^{19}$ GeV. Assuming  $M_D$ to be not too far from the few ${\rm TeV}$ scale, a variety of interesting strong-gravity signals could be expected when the LHC will be turned on a few years from now. 

One particularly dramatic signal \cite{BHF}  would be the production  of  TeV-scale black holes and their characteristic decay by the Hawking process. We expect black holes to be formed with unsuppressed rate provided the impact parameter $b$ of the collision of the (essentially) point-like proton constituents (quarks and gluons) does not exceed the Schwarzschild radius $R_{\rm S}(E_{\rm cm})$ associated with the centre-of-mass energy. This intuition appears to be supported \cite{CTS}  by classical arguments on the formation of a closed trapped surface in the 2-particle collision process.
It was pointed out, however, that the above cross-section estimates have to be amended by taking into account the
finite tranverse size of the colliding objects \cite{KV}. Even in the most optimistic case, by embedding any quantum theory of gravity into string theory, the minimal transverse size of the colliding quanta is given by the string length parameter $l_s$ and the formation of a black hole would  require both $R_{\rm S}(E_{\rm cm}) > b$ {\it and} $R_{\rm S}(E_{\rm cm}) > l_s$. Given  that, in (weakly coupled) string theory $\l_s > l_D$  (where $l_D = M_D^{-1}$ is the true Planck length of the theory), this effect will push the threshold of black hole production to some $E_{\rm th} > M_D$. As a consequence, it is most likely that the energy reach of accelerators such as the LHC will be very marginal for producing semi-classical black holes even in the most optimistic case. This does not mean, however, that no new phenomena will be found below or just near $E_{\rm th}$. Exploring the nature of such exotic phenomena will be the other goal of this work.

The rest of the paper is organized as follows: we first recall, in Section 2,  the ``phase-diagram" of super-Planckian collisions in
superstring theory as it has emerged from work in the late eighties. In Section 3 we present our ansatz for an explicitly unitary S-matrix that is supposedly valid in a large region of our phase diagram. It   embodies the results of previous work as illustrated for several quantities (partly in the main text and partly in an Appendix).
In Section 4  we analyse the final state in the energy window $M_D < E < E_{\rm th}$ and show how the average energy of final particles smoothly {\it decreases}  to the Hawking-temperature value (up to a constant of $O(1)$) as the total energy is {\it increased} towards  $E_{\rm th}$. This suggests the validity, within an energy window, of a sort of antiscaling behaviour for inclusive cross sections that  would provide a smoking gun for the discovery of string/quantum--gravity effects at future colliders. We finally draw, in Section 5, some (partly tentative) lessons that our results could teach on  how information is retrieved when the threshold of black-hole production is crossed.

 \section{Three regimes in super-Planckian collisions}

Collisions of light particles at super-Planckian energies ($E = \sqrt{s} \gg M_D$)\footnote {Here and in the following we shall denote by a subscript $D$ quantities referring to a generic number $D$ of ``large" space-time dimensions, while reserving the label $P$ for $D=4$. We also use units in which $\hbar = c =1$.} have received considerable attention since the late eighties. While in  \cite{tH}  the focus was on $D=4$ collisions in the field theory limit, two groups have carried out  the analysis within superstring theory, possibly allowing for a  number
of ``large" extra dimensions. In the approach due to Gross, Mende and Ooguri (GMO)  \cite{GMO} one starts from a genus-by-genus analysis of fixed angle scattering, and then attempts an all-genus resummation. In the work of Amati, Ciafaloni and Veneziano (ACV) \cite{ACV1,ACV2,ACV3} one starts from an all-order eikonal description of small-angle scattering  and then attempts to push the results towards larger and larger angles.

The picture that has emerged  (see e.g.  \cite{GVPAS} for some reviews) is best explained by working in impact parameter 
($b = 2J/E$) space, rather than in scattering angle ($\theta$) or momentum transfer. We can thus represent the various regimes of super-Planckian collisions
by appealing to an $(E,b)$ plane or, equivalently but more conveniently, to an $(R_S,b)$ plane, where
\beq
\label{RS}
R_S(E) \sim  (G_D E)^{\frac{1}{D-3}} 
\eeq
is the Schwarzschild radius associated with the centre-of-mass energy $E$. Since both coordinates in this plane are lengths, we can also mark on its axes two (process-independent) lengths, the Planck length $l_D$ and the string length $l_s$. We shall use the following definitions:
\beq
\label{def}
l_s = \sqrt{2\alpha' \hbar} = M_s^{-1}  \, , ~ 8 \pi G_D = l_D^{D-2} = M_D^{2-D} ~ ,
\eeq
where $\alpha'$ is the open-string Regge-slope parameter (equal to twice that of the closed string). 
Throughout this paper we will assume string theory to be (very) weakly coupled, so that
\beq
\label{lD}
l_s = (g_s)^{-\frac{2}{D-2}} l_D \gg  l_D \, , \, {\rm i.e.} \, M_D =  M_s (g_s)^{-\frac{2}{D-2}} \gg M_s \, ,
\eeq 
where $g_s$ is the string coupling constant  (our normalization conventions being specified by (\ref{lD})). We shall keep $g_s$ (and hence $l_D/l_s = M_s/M_D$) fixed and very small.

By definition of super-Planckian energy, $R_S > l_D$. Since we also restrict ourselves to $b > l_D$, a small square near the origin is not considered. The rest of the diagram is divided essentially in three regions (see Fig. 1, taken essentially from Refs. \cite{GVPAS}):
\begin{figure}[h] 
\begin{center}
\includegraphics[width=0.7\linewidth]{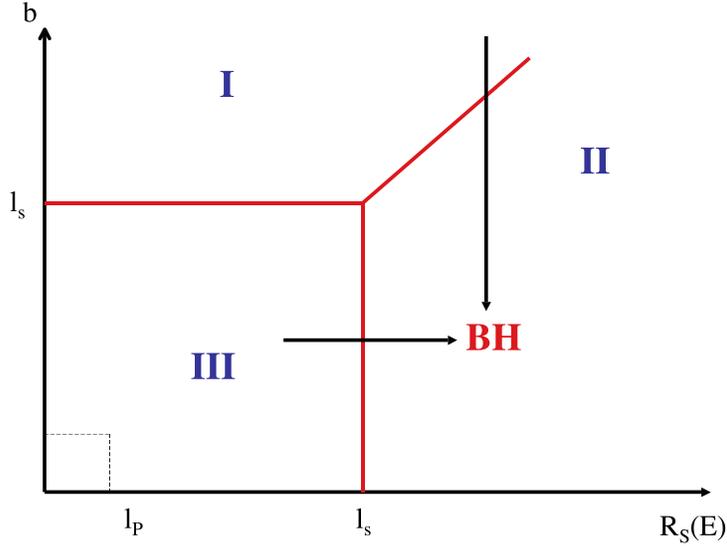}
\end{center}
\vspace{-1.cm} 
\caption{{\smallish Phase diagram of trans-Planckian scattering showing three different kinematic regions and two different paths towards
the regime of large BH production.}} 
\label{fig:FIGURE1} 
\end{figure} 
\begin{itemize}
\item  The first region, characterized by $b > {\rm max}(l_s,R_S)$, is the easiest to analyse and corresponds to small-angle quasi-elastic scattering. 
\item The second region, $R_S > {\rm max}(b, l_s)$, is the most difficult: this is where we expect black-hole formation to show up. Unfortunately, in spite of much effort (see, e.g. \cite{ACV3}, \cite{FPVV}), not much progress has been achieved in the way of going through the $b = R_S > l_s$ boundary of Fig. 1.
\item Finally, the third region ($b, R_S < l_s$), whose very existence depends on working in a string theory framework, has provided some very interesting insight \cite{ACV2} on how string effects may modify classical and quantum gravity expectations once the string scale is approached. The reason why progress could be made by ACV  in this regime, unlike in the previous one, is that string-size effects intervene {\it before} large classical corrections make the problem intractable as $b \rightarrow R_S$.  The physical reason for this is that string-size effects prevent the formation of a putative BH  whose radius would be smaller than $l_s$. Instead, it is found \cite{ACV2} that a maximal deflection angle $\theta_{\rm max} \sim (R_S/l_s)^{D-3}$ is reached at an impact parameter of order $l_s$ (the colliding strings grazing each other). If one considers scattering at angles larger than such $\theta_{\rm max} $, one finds (see e.g. \cite{GVPAS}) an exponential suppression of the cross section, in very good agreement with the results obtained  by GMO \cite{GMO} through their very different approach.
\end{itemize}

In this paper, after recasting the $S$-matrix in a convenient, explicitly unitary form,  we will analyse this third regime further.  We will work at fixed impact parameter and gradually increase the energy to cross the boundary between the second and third regions ($b < l_s = R_S$).
As the boundary is approached,  one arrives at the threshold of BH formation. Indeed, at $\sqrt{s} = E_{\rm th}$,  one expects to form states  that can be seen alternatively  either as strings or as black holes \cite{Corr1, Corr2}. They lie, in an $(M, g_s)$ plane, on the correspondence curve $M= M_s g_s^{-2}$ that separates string states (lying below such curve) from BH states (lying above it). Our aim here is to study details of the scattering process as this threshold is approached from below and compare them with those expected from formation and evaporation of BHs as the same threshold is approached from above.

\section{Explicitly unitary S-matrix in Regions I and III}

Our first claim is that the results of ACV, including their various checks of inelastic unitarity, can be neatly summarized in terms of an explicitly unitary  S-matrix whose validity can be justified well inside Regions I and III. It reads:
\beq
\label{ShatI}
S = {\rm exp}(i \hat{I}) ~,
\eeq
where the hermitian operator $\hat{I}$ is given by:
 \beq
\label{hatI}
  \hat{I} = (\hat{\delta} + \hat{\delta}^{\dagger}) + \sqrt{-2 i (\hat{\delta} - \hat{\delta}^{\dagger})}   (C + C^{\dagger}) =  \hat{I}^{\dagger} ~ ,
\eeq
and the operators $\hat{\delta}$, $\hat{\delta}^{\dagger}$, $C$ and $C^{\dagger}$ satisfy the commutation relations
\beq
\label{CR}
[C,  C^{\dagger} ]  = 1 \; , \;  [\hat{\delta} , \hat{\delta}^{\dagger}] = [C, \hat{\delta}] =  [C, \hat{\delta}^{\dagger}] = 0 ~.
\eeq
Using well-known harmonic-oscillator formulae, Eqs. (\ref{ShatI}), (\ref{hatI}) lead to the more convenient form:
 \beq
\label{ConvS}
 S  =  {\rm e}^{2 i \hat{\delta}}   {\rm e}^{ i \sqrt{-2 i (\hat{\delta} - \hat{\delta}^{\dagger})} ~ C^{\dagger}}  
 {\rm e}^{i \sqrt{-2 i (\hat{\delta} - \hat{\delta}^{\dagger})} ~  C}  ~.
\eeq
The meaning of the $C$, $C^{\dagger}$ operators will be given below, while (with standard notation for normal ordering)
\beq
\label{hatdelta}
\hat{\delta}  = \frac{1}{4 \pi^2}\int_0^{2 \pi} \int_0^{2 \pi} d\sigma_u d \sigma_d :\delta(E, b + \hat{X}_u(\sigma_u) - \hat{X}_d(\sigma_d)): 
~ ,
\eeq
where $\hat{X}_{u,d}$  are hermitian (and commuting) closed-string position operators taken at a fixed value of $\tau$,
and $\delta$ is the tree-level amplitude in impact parameter space:
\beq
\label{delta}
\delta(E,b)  = \frac{1}{(2 \pi)^{D-2}} \int d^{D-2}q  \frac{{\it A}(s,t)}{4s} ~ {\rm e}^{- i q b} ~  , ~~ s = E^2, ~~ t = - q^2 ~ .
\eeq
The explicit form of the tree-level amplitude is:
\beq
\label{Atree}
 \frac{{\it A}(s,t)}{4s}  = (2 \pi G_D) \frac{\Gamma(-\alpha' t/4)}{\Gamma(1+ \alpha' t/4)} \left(\frac{\alpha' s}{4}\right)^{1+ \alpha' t/4}
{\rm e}^{- i \pi \alpha' t/4} ~,
\eeq
and its imaginary part is positive semi-definite, 
\beq
\label{ImAtree}
\frac{{\rm Im} A(s,t)}{4s} =  \frac{2 \pi^2 G_D}{\Gamma^2(1+ \alpha' t/4)}  \left(\frac{\alpha' s}{4}\right)^{1+ \alpha' t/4} \ge 0 \, ,
\eeq
an important feature in order to give meaning
to the definition (\ref{hatI}) of $\hat{I}$. The imaginary part of $\delta$ and thus, from (\ref{hatdelta}),  the anti-hermitian part of $\hat{\delta}$ are given simply by:
\beq
\label{Imdelta}
{\rm Im} \delta = \frac{\pi^2 G_D~ s~ l_s^2}{4 (\sqrt{\pi} Y \l_s)^{D-2}} {\rm e} ^{-b^2/l_s^2 Y^2} ~ ,
\eeq
where\footnote  {Note a change of notation here with respect to ACV. In later expressions the argument of the log is not necessarily the total energy, and thus we will occasionally assimilate $Y$ to a numerical constant.}  $Y = \sqrt{{\rm log}(\frac{\alpha' s}{4})}~$.

Equation (\ref{Imdelta})  introduces a characteristic impact parameter $b_I = l_s Y$ above which $\delta$ is basically real and thus
$\hat{\delta}$ is hermitian. We will be mainly concerned with the opposite region, $b \le b_I$, but let us first recall some properties of the $S$-matrix in the former regime.
At $b \gg b_I$, the behaviour of $\delta$ is dominated by the pole at $t=-q^2 =0$ in $A(s,t)$ with the result:
\beq
\label{delargeb}
\delta = \left(\frac{ b_c}{b}\right)^{D-4} \equiv \frac{G_D s}{\Omega_{D-4}} b^{4-D} ~ ,
\eeq
with $\Omega_d = \frac{2 \pi^{d/2}}{\Gamma(d/2)}$ the solid angle in $d$ dimensions. As is well known \cite{ACV1}, the above formula reproduces, at very small angle, the classical relation between impact parameter and (centre-of-mass) scattering angle in the (Aichelbourg--Sexl) metric of a relativistic point-like source:
\beq
\label{AS}
\theta = \frac{8 \pi G_D \sqrt{s}}{\Omega_{D-2} b^{D-3}} ~ .
\eeq

As pointed out by ACV, however, the passage from  $\delta$ to
$\hat{\delta}$ introduces another impact parameter scale $b_{\rm DE}$ below which the elastic amplitude is suppressed in favour of channels where the initial particles are diffractively excited (with a slight abuse of the word ``diffractive") through graviton exchange. One finds that the elastic amplitude is suppressed as 
\beq
\label{DE}
|A_{\rm el}| \sim {\rm e}^{-2 \delta_{\rm DE}} ~~, ~~  \delta_{\rm DE} \equiv \frac{\pi^2 (D-3) G_D s l_s^2}{2 b^{D-2} \Omega_{D-2}}~.
\eeq
This defines the critical $b$ by the condition that the exponent is 1 for $|A_{\rm el}|^2$, i.e.
\beq
\label{bDE}
b_{\rm DE}^{D-2} = \frac{2\pi^2 (D-3) G_D s l_s^2}{\Omega_{D-2}}~,
\eeq
where we have to keep in mind that this is only valid if $b > b_I$. Notice that $b_{\rm DE} = b_I$ for $s = M_s^2/g_s^2 \equiv M_*^2$. This new scale $M_*$  plays an important role in the following discussion. It is basically the geometric mean between the string scale $M_s$ and $E_{\rm th}$, but, while   $M_* = M_{\rm P}$ in $D=4$, $M_*$ is 
parametrically larger than $M_D$ for $D>4$. A simple explanation of its dynamical origin is proposed  in Section 5.
The situation is sketched in Figs. 2 and 3 for $D=4$ and $D=6$, respectively.
\begin{figure}[h] 
\begin{center}
\includegraphics[width=0.7\linewidth]{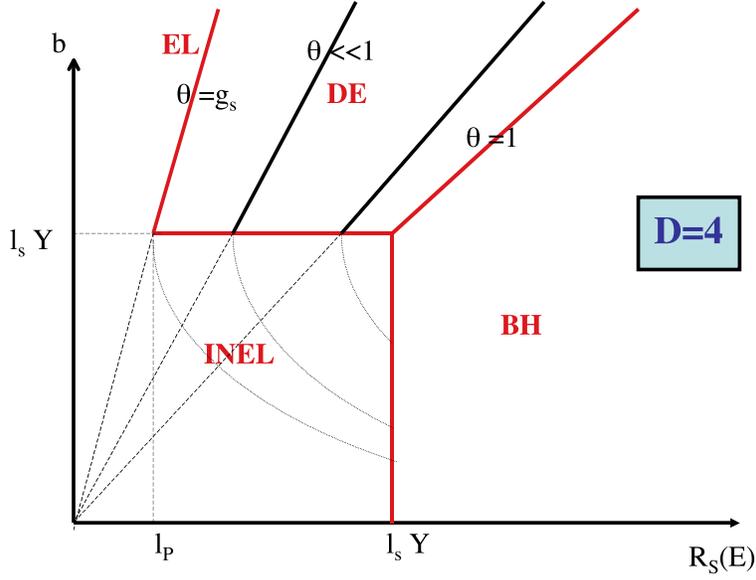}
\end{center}
\vspace{-1.cm} 
\caption{{\smallish Further subdivision of the phase diagram of Fig. 1 for $D=4$.}} 
\label{fig:FIGURE1} 
\end{figure} 
\begin{figure}[h] 
\begin{center}
\includegraphics[width=0.7\linewidth]{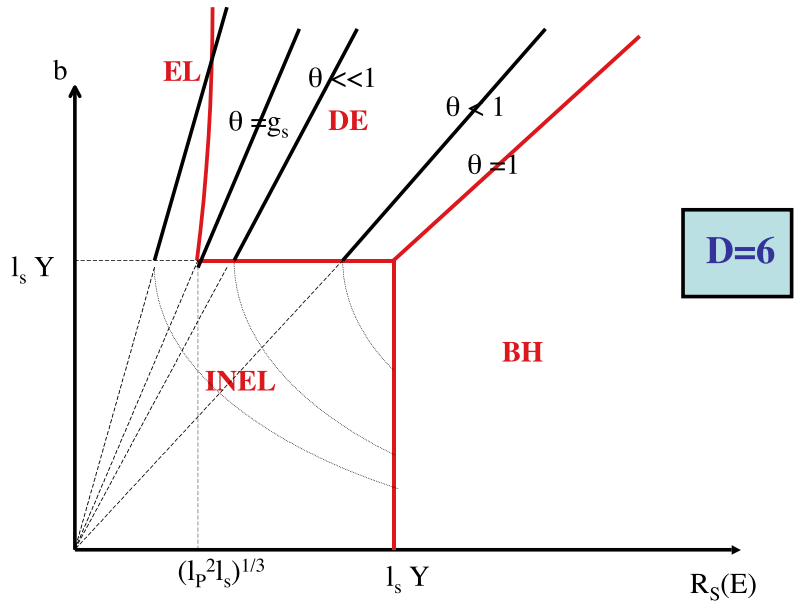}
\end{center}
\vspace{-1.cm} 
\caption{{\smallish Further subdivision of the phase diagram of Fig. 1 for $D=6$. A qualitatively similar behaviour holds for any $D>4$.}} 
\label{fig:FIGURE1} 
\end{figure} 
The curve (\ref{bDE}) divides Region I in two parts: in the upper one, elastic scattering dominates (and elastic unitarity is  satisfied); in the lower part, inelastic scattering dominates (and the $S$-matrix explicitly satisfies inelastic unitarity). Of course we lose control over our approximations if we move towards the region $b \sim R_S> l_s$, the large-angle regime. For some attempts to tackle this region see, e.g., \cite{ACV3}, \cite{FPVV}. 

Let us now describe the region $b < b_I$.
Here (modulo an irrelevant $b$-independent term) the real part of $\delta$ behaves as:
\beq
\label{smallbdelta}
{\rm Re}\delta = - \frac{G_D~ s~b^2}{(D-2)(\sqrt{\pi} Y l_s)^{D-2}}~.
\eeq

If we were to neglect all non-elastic effects this phase shift would produce \cite{ACV1} the same relation between scattering angle and
impact parameter that results from the geodesic motion of a particle in the metric of a homogeneous finite-size relativistic beam \cite{hombeam}:
\beq
\label{smallbdef}
\theta = \frac{16 \pi G_D ~\rho~ b}{D-2} ~~, ~~ \rho = \frac{E}{(\sqrt{\pi} Y l_s)^{D-2}}~,
\eeq
where $Y l_s $ plays the role of  the transverse size of the beam. The curved lines in Figs. 2 and 3, corresponding to fixed $\theta$, are sketching such a behaviour.

Inelasticity now comes from two distinct sources:
\begin{itemize}
\item{1.} From replacing ${\rm Re}\delta$ by $\frac{1}{2}  (\hat{\delta} + \hat{\delta}^{\dagger})$ in order to take account of DE (as we did in Region I). 
This effect leads, as in Region I,  to an exponential damping of $|A_{\rm el}|$.  However, instead of (\ref{DE}) one finds:
\beq
\label{bDEI}
\delta_{\rm DE} = \frac{\pi^2 G_D s l_s^2}{4 (l_s \sqrt{\pi} Y)^{D-2}} ~, ~~~b < b_I ~ .
\eeq
It is easy to check that  (\ref{bDE}) approaches  (\ref{bDEI})  for $b \rightarrow l_s Y$, hence the behaviour  (\ref{DE}) saturates at $b \sim b_I$.

\item{2.} From the fact that
$\delta$ has now acquired an imaginary part given by (\ref{Imdelta}).  This is related to the possibility of ``cutting" the exchanged gravi-reggeons, i.e. to the fact that, in string theory, because of (old Dolen--Horn--Schmit) duality, gravi-reggeon exchange in the $t$-channel is dual to some $s$-channel intermediate states. For open-string collisions these intermediate states correspond to a pair of excited open strings, while for closed-string scattering they represent a single excited  closed string. String breaking will make these states yield showers of light particles.
The operators $C, C^{\dagger}$ are just destruction and creation operators for a cut gravi-reggeon (CGR), i.e. for one such kind of final state. More generally, they should carry a label corresponding to
the energy-momentum (and possible internal quantum numbers) of the CGR.
\end{itemize}
When a diagram with $n$ exchanged gravi-reggeons is considered, the full imaginary part gets contributions from different final states, corresponding to cutting any number $m \le n$ of gravi-reggeons. The relative weights of these contributions
were found long ago by Gribov, Abramovskii and Kancheli \cite{AGK} and are known as the AGK rules.
It is quite remarkable that precisely the AGK rules allow us to rewrite the full S-matrix in the simple form (\ref{ConvS}) (see the appendix for a sketch of the proof).
It is also  amusing to realize that exactly the same damping of $|A_{\rm el}|$  as (\ref{bDEI})  follows from the imaginary part of $\delta$ via eq. (\ref{Imdelta}) when $b < b_I$.  We shall come back to other such similarities in the following section.

\section{Properties of the final state in the energy window}
Going back to  Figs. 2 and 3, let us emphasize again that Region I of Fig. 1 gets further divided in two parts by the line given by eq. (\ref{bDE}). This line ends
at $b=b_I$ and $E = M_* = M_s/g_s$, where it joins the horizontal segment $b=b_I, M_* < E < E_{\rm th} = M_s/g_s^2$.
Below such a segment CGRs are copiously produced. In order to understand this qualitatively,
let us consider the generating function
\beq
\label{GF}
{\rm exp}(W(z)) \equiv    \langle in| S^{\dagger} z^{N_{\rm CGR} } S |in \rangle  \, , ~~  N_{\rm CGR} \equiv  C^{\dagger} C ~,
\eeq
where $N_{\rm CGR} $ counts the number of ``cut gravi-reggeons".
Evaluating $W(z)$ is now a simple exercise in harmonic oscillator operators,  simplified by the well-known identity:
\beq
\label{Ident}
z^{N_{\rm cgr} } = : {\rm exp}\left((z-1) C^{\dagger} C\right) : ~.
\eeq
Using properties of the coherent states one arrives at the simple result:
\beq
\label{W}
W(z) =  4 (z-1)  {\rm Im} \delta  ~ ,
\eeq
which implies that the distribution of $N_{\rm CGR}$ is exactly poissonian, with an average 
given by:
\beq
\label{Nav}
\langle N _{\rm CGR}\rangle =  4 {\rm Im} \delta = \frac{\pi^2 G_D~ s~ l_s^2}{(\sqrt{\pi} Y \l_s)^{D-2}}~ 
= O\left(\frac{s}{M_*^2}\right) ~.
\eeq

Because of the operators $\hat{X}$ entering in $\delta$, the cut gravi-reggeons do not exhaust, however, the final state.
As explained in the previous section, their effect is to produce diffractive excitation of the initial particles through graviton exchange, i.e. something that preserves the initial state's conserved quantum numbers. If we insist that no such excitation take place  the cross section is damped by a factor that, as noticed in Section 3, is just ${\rm exp}(- 4 {\rm Im} \delta)$.
If we also forbid particle production via CGRs we are penalized again by the same exponential factor (as is made clear by taking the limit $z \rightarrow 0$ in Eqs. (\ref{GF}) and (\ref{W})). In conclusion, the elastic cross section is damped as:
\beq
\label{Eldamp}
\sigma_{\rm el} \sim {\rm exp}(- 8 {\rm Im} \delta) = {\rm exp}\left[- \frac{2 \pi^2 G_D~s~l_s^2}{(\sqrt{\pi} Y l_s)^{D-2}} \right] ~.
\eeq
Amusingly, such a suppression is quite similar to a factor ${\rm exp}(- S_{\rm bh})$ where $S_{\rm bh}$ is the Bekenstein--Hawking entropy of a black hole of mass $\sqrt{s}$. Such a suppression is known to appear in the elastic amplitude
for the scattering of a particle off a classical BH when the impact parameter goes below that of classical capture (e.g. \cite{NS}). For $D=4$ the agreement holds up to a numerical factor $\frac{1}{2Y^2}$, while for $D>4$ the functional dependence is different. For any value of $D$, however, agreement with
a BH-type suppression holds (up to factors of $O(1)$) as one approaches $E_{\rm th}$.

Let us now look instead at the typical final state that roughly saturates the cross section. To begin with,  the average mass of the DE states was computed in \cite{ACV2} as:
\beq
\label{MDE}
\langle M \rangle_{\rm DE} \sim \sqrt{s}  \left( \frac{R_S}{l_s} \right)^{D-3}Y^{2-D}~ .
\eeq
In words, $\langle M \rangle_{\rm DE} $ increases as the energy is increased in the window until it gets close to a small fraction of the total energy
at $E = E_{\rm th}$.  In any case, even when they are much lighter than $\sqrt{s}$, DE states tend to be fast and thus  take out a finite fraction $f$ of the total incoming energy.

The rest of the initial energy will be shared, in the final state, between $N_{\rm CGR}$ CGRs, giving an average energy per CGR:
\beq
\label{ECGR}
\langle E \rangle _{\rm CGR} = (1-f) \frac{\sqrt{s}}{\langle N_{\rm CGR} \rangle}  \sim (1-f) M_s Y^{D-2} \left( \frac{l_s}{R_S} \right)^{D-3}  ~.
\eeq

This time, the average energy of the final states {\it decreases} as the energy is {\it increased} in the window. Once more, for $D=4$
the functional dependence on $\sqrt{s}$ of $\langle E \rangle_{\rm CGR}$ follows that of a thermal stectrum with a Hawking temperature 
$T_H \sim \hbar/R_S$, while for $D>4$ the functional dependence is different\footnote {Similar differences between $D=4$ and $D>4$ in the string--BH correspondence were already noted in \cite{Corr2}.}. For all values of $D$, however, agreement (up to factors $O(1)$) with Hawking-temperature expectations occurs at $E_{\rm th}$. An ``antiscaling"  behaviour, 
corresponding to 
\beq
\label{antisc}
\langle E \rangle_{\rm CGR} \sqrt{s} = M_*^2 = M_s^2 g_s^{-2} ~,
\eeq
holds for any $D$ in the energy window $M_* < E < E_{\rm th}$. It is easy to guess the generic form
 that such an antiscaling behaviour should take at the level  of inclusive cross sections for the final particles after CGR decay. For instance, the single-particle spectrum will be given as:
 \beq
 \label{singlep}
 \frac{\omega}  {\sigma}  \frac{d \sigma}  {d\omega} = \frac{x}  {\sigma}  \frac{d \sigma}  {dx} = \left( \frac{s}{M_*^2}\right)~ f(x)  ~ ,
 \eeq
where $\omega$ is the energy of the final particle and we have introduced the (anti)scaling variable: 
\beq
\label{x}
x  =  \frac{\omega}{T_{\rm eff}}~~,~ T_{\rm eff} \equiv  \frac {M_*^2}{E} = \frac {M_s^2}{g_s^2E} ~.
\eeq
Energy conservation imposes the sum rule:
\beq
\label{Econs}
1 =  \int dx f(x) \, ,
\eeq
while the average multiplicity is given  by
\beq
\label{mult}
\langle n \rangle =  \frac{s}{M_*^2}  \int \frac {dx}{x}  f(x) ~ = O\left( \frac{s}{M_*^2}\right),
\eeq
in accordance with (\ref{Nav}). A  typical form of $f(x)$, taking into account both phase space and a high-frequency cutoff, would be  $f(x) = c x^{D-2} {\rm exp} (-c' x)$.

 Generalization to multiparticle inclusive cross sections is straightforward. Keeping in mind that CGRs are uncorrelated, one can expect the generating functional of multiparticle correlations ($W= {\rm log} Z$) to have the weak-correlation form:
 \beq
 \label{genfnctnl}
 W[z(\omega)] = \frac{s}{M_*^2} \sum_{n =1}^{\infty}  \frac{1}{n!} \int  \rho_n(x_1, \dots, x_n) \prod_{i=1}^{n} d\omega_i (z(\omega_i) -1)  \, , ~ x_i = \frac{\omega_i}{T_{\rm eff}} ~, 
 \eeq
where some correlations must be present because of energy conservation.
\begin{figure}[h] 
\begin{center}
\includegraphics[width=0.7\linewidth]{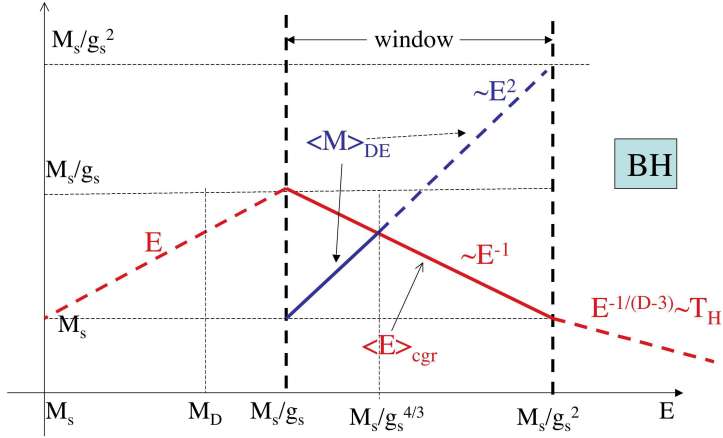}
\end{center}
\vspace{-1.cm} 
\caption{{\smallish Energy dependence of $\langle M \rangle_{\rm DE} $ and $\langle E \rangle_{\rm CGR}$ in the energy window $M_s/g_s < E < M_s/g_s^2$ showing in particular the anti-scaling behaviour of the latter. The two curves cross at the energy $M_s g_s^{-4/3}$ corresponding to closing the rapidity gap between the two DE states.}}
\label{fig:FIGURE1} 
\end{figure} 
In Fig. 4 we show the dependence of $\langle M \rangle_{\rm DE} $ and $\langle E \rangle_{\rm CGR}$ in the above energy window. Note the  opposite qualitative behaviour of $\langle E \rangle_{\rm CGR}$ and  $\langle M \rangle_{\rm DE} $. The former decreases from  $M_*$  to $M_s$ while the latter increases from $M_s$ to a fraction of the total energy at $E_{\rm th}$.  The two curves intersect
when $E \sim M_s g_s^{-4/3}$, i.e. when $\langle M \rangle_{\rm DE} ^2 \sim M_s E$.
At this energy, after string break-up,  the two
DE states will produce two jets of particles at some (finite) angle relative to the ``beam axis", but with limited transverse momenta ($p_t < M_s$)
relative to the jet axis itself. It is  clear that,  when  $\langle M \rangle_{\rm DE} ^2 \ge M_s E$,  no rapidity gap will be left between the final particles of the two jets\footnote{I am grateful to M. Ciafaloni for
an interesting exchange on this issue.}, making the distinction between DE and  CGR states practically impossible (for this reason the solid line for $\langle M \rangle_{\rm DE} $ is not continued above $E \sim M_s g_s^{-4/3}$).
 All that matters, from there on, is the number of handles  that are cut: each one of them corresponds to an excited closed string produced by the collision. Our $S$-matrix gives the final state, $|f\rangle = S |2\rangle$,  as a coherent state of CGRs with a very large average occupation number. 

 The characteristic features of the final state would represent a clear signal of new string or quantum-gravity physics, already well below the threshold of BH production, provided the higher-dimensional model has $M_s \ll M_D$. In the following section we will argue that these features may also suggest the way coherence is kept even above such a threshold.

\section{Implications for the information paradox }

Which are the possible implications of our results when one crosses the threshold for BH production?
At the moment, we can only put forward some educated guesses:
\begin{itemize}
\item DE should carry away the energy that, even classically, goes to future infinity even when a BH is formed. It carries with it a fraction of the incoming energy but also the (nominally) conserved global quantum numbers of the initial state and could thus play an important role in guaranteeing their actual conservation. However, in our calculation, DE depends on $l_s$ while in the end it should not. Possibly $l_s$-dependence should freeze when $R_s > l_s$.
\item The CGRs will start interacting with each other (see e.g. the so-called H-diagram of \cite{ACV3}), showers will develop, and the original Poisson distribution of the CGR should turn into an approximately thermal one (with BE or FD statistics) for the actual final particles.
\item The eikonal phase $\delta$ will pick up an imaginary part over the one due to CGR production, particularly as we approach BH production from Region I. It is amusing to speculate on how the expected result at a point well inside Region II can be obtained
by approaching it from Regions I and III. In Region I the general interpretation of the phase shift is in terms of some
classical (retarded vs advance)-time delay \cite{FPVV}. When a critical $b \sim R_s$ is reached, the time delay diverges
 (for any $D$) like $R_s {\rm log} (b-R_S)$. As we go below this critical value of $b$, the time delay develops an  imaginary part $\sim i \pi R_S$ (since there is no classical trajectory going off to future infinity); when inserted in the phase $i E\Delta t$, this gives:
\beq
\label{AelRI}
|A_{\rm el}| \sim {\rm exp}(- \pi E R_s) \, , 
\eeq
namely the correct damping (up to numerical factors $O(1)$) for all $D$.
How does this match the behaviour obtained by crossing the boundary horizontally rather than vertically?
The answer is, again, that the $l_s$ dependence in Eq. (\ref{Eldamp}) should saturate when $l_s$ approaches $R_s$, so that:
\beq
\label{AelRIII}
|A_{\rm el}| \sim  {\rm exp} \left[-  G_D~s~l_s^{ 4-D} \right] \rightarrow  {\rm exp}\left[-  G_D~s~R_s^{4-D} \right]  \sim  {\rm exp}(- E R_s) ~,
\eeq
in agreement with (\ref{AelRI}). The same correspondence with BH physics would come out at the level of inclusive spectra if  the $l_s$ dependence in Eq. (\ref{singlep}) saturates at $R_S = l_s$, giving:
\beq
\label{sat}
\frac{s}{M_*^2} \rightarrow E R_S \sim S_{\rm BH} ~ , ~ T_{\rm eff}  \rightarrow 1/R_S  \sim T_{H} ~ , ~ x \rightarrow \omega/T_{H} ~ .
\eeq
\end{itemize}

It is interesting to try to understand the origin of the new scale $M_*$, which plays such an important r\^ole in the energy window. At the semi-classical level at which we are working physics should depend basically on $S_{\rm cl}/\hbar$, where  $S_{\rm cl}$ is the action evaluated on the classical solution (see \cite{FPVV} for a more precise discussion). In the presence of external sources providing a non-trivial
energy-momentum tensor $T_{\mu \nu}$,  $S_{\rm cl}/\hbar$ scales as:
\beq
\label{Sscaling}
 S_{\rm cl}/\hbar \sim l_{\rm P}^{2-D} \int d^D x (G_D T) \partial^{-2} (G_D T) ~ ,
 \eeq
 a pure number, as it should. $T$ is proportional to the total energy $E$. However, while
 much above $E_{\rm th}$ the length scales entering (\ref{Sscaling}) are all related to $R_S$, in the window they are all dominated by $l_s$. This is why 
 (\ref{Sscaling}) scales differently in the two regimes:
 \bea
\label{DSscaling}
 S_{\rm cl}/\hbar \sim l_{\rm P}^{2-D} (GE)^2 R_S^{4-D} ~ = \left(\frac{R_S}{ l_{\rm P}}\right)^{D-2} ~ ~~&,& E >  E_{\rm th} = M_s g_s^{-2} \nonumber \\
  S_{\rm cl}/\hbar \sim l_{\rm P}^{2-D} (GE)^2 l_s^{4-D} ~\sim g_s^2 \frac{E^2}{M_s^2}  =
   \frac{E^2}{M_*^2} ~~~~~~~~~~& ,&M_*= M_s g_s^{-1} <E < E_{\rm th} ~.
   \eea
 The  two behaviours given in (\ref{DSscaling}) agree only for $D=4$ and otherwise provide an explanation for the different energy dependences of $T_H$ and $T_{\rm eff}$.
 
Let us conclude by summarizing the  main points:
\begin{itemize}
\item we have been able to recast the main results of \cite{ACV1,ACV2,ACV3} in the form of an approximate, but  exactly unitary,  S-matrix whose range of validity covers a large region of the kinematic energy--angular-momentum plane;
\item we have studied the nature of the dominant final states in a window of energy and impact parameter at whose boudary we expect black-hole formation  to begin;
\item we have found there a sort of precocious black-hole behaviour, in particular an ``anti-scaling" dependence of the average energy of the final particles from the initial energy, quite reminiscent of the inverse relation between black-hole mass and temperature;
\item this antiscaling behaviour introduces, through the  variable $x = \omega E/ M_*^2$, a new energy scale $M_* = M_s g_s^{-1}$, whose physical origin we have traced back to how the semi-classical action scales in a regime dominated by string-size effects;
\item we believe that these results have a twofold application: a conceptual one in the search for an 
explicit resolution of the information paradox, and a more phenomenological one in the context of the expected string/quantum-gravity signals at collider physics, which are expected in models with large extra dimensions.
\end{itemize}

Our explicit results confirm, to a large extent, what had already been guessed at a more qualitative level, by many authors, including those of Refs.  \cite{ACV1,FPVV, BF},  and hopefully set a framework for starting a more quantitative study of super-Planckian collisions up to, and perhaps even above, the BH-production threshold.

\section*{Acknowledgements}
It is a pleasure to thank D. Amati, M. Ciafaloni, G.T. Horowitz, E. Rabinovici, J. G. Russo and A. Schwimmer for  interesting discussions.

\section*{Appendix}

We will show here that applying the AGK rules gives exactly the result presented in the main text.

We want to identify the inelastic channels that saturate unitarity when $\hat{\delta}$ is not Hermitian, i.e. when
${\rm Im} \delta \ne 0$. Let us neglect DE and thus simply set $\hat{\delta}= \delta$ in our expression (\ref{ConvS}) for the S-matrix.  Recalling that:
\begin{itemize}
\item  ${\rm Im} \delta \ne 0$ comes from the imaginary part of the single (Reggeized) graviton exchange;
\item The leading-eikonal formula $S = {\rm exp}  (2 i \delta)$  corresponds to summing over an infinite set of diagrams  corresponding to the exchange of an increasing number $n =0, 1, 2, \dots$ of gravitons;
\item The rules for reconstructing the imaginary part of a diagram with $n$ (positive signature) exchanged Reggeons
as a sum of contributions coming from ``cutting" (i.e. taking the imaginary part of) $m$ of them were given long ago by Abramowski, Gribov and Kancheli \cite{AGK} (the so-called AGK rules) and were later justified by several authors (e.g. \cite{AGKJ}).
\end{itemize}

These rules (when directly applied in impact parameter space) tells us that the full imaginary part of the $n$-graviton exchange graph is the result of a contribution
\beq
\sigma_m^n = (-1)^{n-m}   \frac{(4{\rm Im} \delta )^n}{m! (n-m)!} ~, ~ n=1, 2, \dots ~,~ m = 1, 2, \dots \le n
\eeq
due to cutting $m$ out of $n$ gravi-reggeons, and a contribution
\beq
\sigma_0^n = (-1)^{n}    \frac{(4{\rm Im} \delta)^n}{n! } - 2 {\rm Re} S^{(n)} ~, ~ n=1, 2, \dots ~ ,
\eeq
when no gravi-reggeon is cut, where $S^{(n)} $ is the full $n$-GR exchange contribution to the S-matrix.
 As the symbol $\sigma$ indicates, these are also to be interpreted as contributions to 
cross sections into inelastic channels corresponding to $m$ cut gravi-reggeons.
It is trivial to check that, for any given $n$, the sum of all contributions from $m=0$ to $m=n$ gives back twice the full imaginary part of $T^{(n)} \equiv i(1-S^{(n)}) $, as it should.

Let us now construct, as in the main text, the generating function of   many CGR cross sections:
\beq
\label{GFdef}
F(z) = {\rm exp} (W(z)) =  \sum_{m=0}^{\infty} z^m \sigma_m = \sigma_0 + \sum_{m=1}^{\infty} z^m \sum_{n=m}^{\infty} \sigma_m^n ~ .
\eeq
This expression is easily computed\footnote{The only slightly subtle point here consists in recovering the elastic
cross section from the AGK rules: this needs the explicit inclusion of the the trivial contribution to the $S$-matrix, $S=1$. }  to give:
\beq
\label{GFres}
F(z) = {\rm exp} \left(4 (z-1)  {\rm Im} \delta \right) ~ ,
\eeq
i.e. the result (\ref{W}) for $W(z)$.

\end{document}